\documentclass[10pt,conference]{IEEEtran}
\usepackage{cite}
\usepackage{amsmath,amssymb,amsfonts}
\usepackage{algorithmic}
\usepackage{graphicx}
\usepackage{textcomp}
\usepackage{xcolor}
\usepackage{booktabs}
\usepackage{tikz}

\usepackage[printwatermark]{xwatermark}
\newwatermark[allpages,color=gray!50,angle=45,scale=2,xpos=0,ypos=0]{Preprint}

\usepackage{macros}

\begin{document}

\title{Automated Black-box Testing of Mass Assignment Vulnerabilities in RESTful APIs}

\author{%
	\IEEEauthorblockN{%
		Davide Corradini\IEEEauthorrefmark{1}, 
		Michele Pasqua\IEEEauthorrefmark{3} and 
		Mariano Ceccato\IEEEauthorrefmark{4}}
	\IEEEauthorblockA{%
		\textit{Department of Computer Science} \\
		\textit{University of Verona} -- Verona, Italy \\
		Email: %
		\IEEEauthorrefmark{1}davide.corradini@univr.it, 
		\IEEEauthorrefmark{3}michele.pasqua@univr.it, 
		\IEEEauthorrefmark{4}mariano.ceccato@univr.it}
}

\makeatletter
\let\@oldmaketitle\@maketitle
\renewcommand{\@maketitle}{\@oldmaketitle
	\medskip
	\begin{center}
	\begin{tikzpicture}
	\node (A) at (0,1.25) {Paper accepted for publication in the proceedings of:};
	\node at (0,0.75) {\emph{45$^\text{th}$ IEEE/ACM International Conference on Software Engineering (ICSE 2023)}};
	\node (B) at (0,0) {\footnotesize The present document is the preliminary version of the work prior to peer-review. The final version can be found on the publisher website.};
	\node (C) at (0,-.75) {\footnotesize \underline{How to cite this paper}:};
	\node (D) at (0,-1.25) {\footnotesize D. Corradini, M. Pasqua and M. Ceccato, ``Automated Black-box Testing of Mass Assignment Vulnerabilities in RESTful APIs'',};
	\node (E) at (0,-1.6) {\footnotesize 2023 IEEE/ACM 45$^\text{th}$ International Conference on Software Engineering (ICSE), 2023.};
	\filldraw[rounded corners=2pt,fill=gray,draw=gray!25,opacity=0.25] (B.west |- 52, 52 |- E.south) rectangle (B.east |- 52, 52 |- A.north);
	\end{tikzpicture}
	\end{center}}
\makeatother

\maketitle

\bigskip

\begin{abstract}
%
%
%
%
%
%
%
%
%
Mass assignment is one of the most prominent vulnerabilities in RESTful APIs. This vulnerability
originates from a misconfiguration in common web frameworks, such that naming convention and automatic binding can be exploited by an attacker to craft malicious requests writing confidential resources and (massively) overriding data, that should be read-only and/or confidential. 

In this paper, we adopt a black-box testing perspective to automatically detect mass assignment vulnerabilities in RESTful APIs. Execution scenarios are generated purely based on the OpenAPI specification, that lists the available operations and their message format.
Clustering is used to group similar operations and reveal read-only fields, the latter are candidate for mass assignment. Then, interaction sequences are automatically generated by instantiating abstract testing templates, trying to exploit the potential vulnerabilities. Finally, test cases are run, and their execution is assessed by a specific oracle, in order to reveal whether the vulnerability could be successfully exploited.

The proposed novel approach has been implemented 
and evaluated on a set of case studies written in different programming languages. The evaluation highlights that the approach is quite effective in detecting seeded vulnerabilities, with a remarkably high accuracy. 
\end{abstract}


\section{Introduction}
\label{sec:intro}

RESTful APIs (or REST APIs for short) are becoming the standard technology to access web-oriented resources and to interconnect software systems across the public Internet. They operate on the web using the HTTP protocol, typically exchanging JSON payloads, and for this reason they are also known as {\em Web APIs}.


Considering their dominant responsibility as cornerstone integration technology to interconnect different computer systems, it is crucial to reveal defects and vulnerabilities in their implementation as soon as possible. In fact, the security of the overall integrated system builds on top of the security and reliability of the atomic REST APIs that take part to the composition.

Specific vulnerabilities are known to impend REST APIs, as reported by the OWASP Foundation in their annual survey\footnote{\url{https://owasp.org/www-project-api-security/}}. In particular, {\em mass assignment} is among the most prominent vulnerabilities that are peculiar to this programming domain. This vulnerability originates from a wrong configuration of common REST API frameworks, that typically provide automatic binding between input data fields (controlled by a potential attacker) to internal data representation (e.g., to database columns). A successful exploit to a mass assignment vulnerability would allow attackers to manipulate private data, provided that they are able to guess the names in the internal data structures (e.g., database tables and columns) used by the REST API. 

A largely adopted perspective when automatically testing REST APIs~\cite{Corradini2022,Atlidakis2020SecurityProperties,Mai2020MetamorphicSecurityTesting,Yang2017RestSep,Yang2016RestPL} is black-box testing, that does not require access to the APIs source code. Also in this work we adopt a black-box approach, since it allows to successfully test REST APIs that are implemented in different languages, that are based on any possible REST framework, and that are possibly composed by closed-source components. Indeed, a black-box perspective allows to test REST APIs independently of their internal architecture (typically micro-service oriented).

In this paper, we propose a novel approach to automatically test REST APIs with respect to mass assignment vulnerabilities. The specification of the REST API to test is analyzed to identify what data should not be overwritten by incoming requests, namely data that are supposed to be read-only and hence candidate for mass assignment attacks. Our analysis first identifies groups of operations with similar data using clustering, then some heuristics are exploited to identify read-only fields. Subsequently, concrete test scenarios are automatically generated by instantiating a catalog of abstract test templates. Eventually, a security oracle monitors the execution of these test cases to reveal when a vulnerability is exposed. 

Empirical assessment suggests that our approach is very accurate in detecting mass assignment vulnerabilities, in fact all the vulnerabilities could be detected in almost all the considered case studies, with no false positive.

To the best of our knowledge, no automated black-box approach is available in the literature to detect mass assignment vulnerabilities on REST APIs.

The paper is organized as follows. After covering the required background on REST APIs and mass assignment in Section~\ref{sec:background}, Section~\ref{sec:approach} presents our novel testing approach for mass assignment. Section~\ref{sec:settings} presents the definition of our experimental design, while Section~\ref{sec:results} applies it and comments the collected results. After comparing our approach with related work in Section~\ref{sec:related}, Section~\ref{sec:conclusion} closes the paper.


\section{Background}
\label{sec:background}

\subsection{RESTful APIs}

A RESTful API (or REST API for short) is a web API that adheres to the REST (REpresentational State Transfer) architectural style~\cite{Fielding2000ArchitecturalStyle}. 
REST APIs provide a uniform interface to create ($C$), read ($R$), update ($U$), and delete ($D$) resources (known as CRUD semantics). A resource is identified by a HTTP URI, and CRUD operations are typically mapped to the HTTP methods \texttt{POST}, \texttt{GET}, \texttt{PUT} (or \texttt{PATCH}) and \texttt{DELETE}.


As an example, consider \emph{VAmPI}\footnote{\url{https://github.com/erev0s/VAmPI}}, an open source REST API to manage users and books of a library. The HTTP URI pointing to a user resource is \texttt{/users}, and the HTTP operations \texttt{GET /users} and \texttt{POST /users/register} are used to retrieve the list of registered users and to register a new user in the system, respectively.

A REST API may accept input parameters to specify additional information for executing an operation, such as the identifier of the user to retrieve (\eg \texttt{/users/\{username\}}) or a structured object to be registered in the system in the body of the request by means of a \texttt{POST} method.

\subsection{The \openapi Specification}

\begin{figure}[t]
\begin{lstlisting}[language=yaml,numbers=none,xleftmargin=3pt,xrightmargin=3pt]
/users:
  get:
    operationId: retrieve_all_users
    responses:
      '200':
        description: See all details of the users
        content:
          application/json:
            schema:
              type: array
              items:
                type: object
                properties:
                  admin:
                    type: boolean
                  email:
                    type: string
                  password:
                    type: string
                  username:
                    type: string
/users/register:
  post:
    operationId: register_new_user
    requestBody:
      content:
        application/json:
          required: true
          schema:
            type: object
            properties:
              username:
                type: string
              password:
                type: string
              email:
                type: string
\end{lstlisting}
\vspace*{-5pt}
\caption{Part of the OpenAPI specification for \emph{VAmPI}.}\label{code:openapiexample}
\end{figure}


\openapi\footnote{\url{https://www.openapis.org/}} defines a standard to document REST APIs. According to \openapi, an API service is described using a structured file (either YAML or JSON) that specifies how to reach the API using a URI, which authentication schema is adopted, and the details of all the operations available in the API. In particular, for each operation the input parameters (and their schema) to be used in requests, and the schema of responses are given. 


Fig.~\ref{code:openapiexample} contains an excerpt of the \openapi specification for our example \emph{VAmPI}. After an initial header that specifies versions, licenses, and the base URL of the API (not shown in the code for space reasons), the \openapi specification contains an array of \emph{paths}, namely the list of URL paths available in the API. 
In our example, we just focus on two URL paths: \texttt{/users}, and \texttt{/users/register}. Each path supports one or more HTTP methods. A path together with an HTTP method composes an \textit{operation}, which is usually identified by an \emph{operation ID}. For instance, the method \texttt{GET} in \texttt{/users} (\textit{retrieve all users}) is used to retrieve the list of all registered users in the system, while the method \texttt{POST} in \texttt{/users/register} refers to the operation \textit{register user}, meant to register a new user in the system. 


Input and output of operations are associated to a \emph{schema} that specifies their type and, optionally, a set of constraints on values (\eg a \emph{minimum} or \emph{maximum} value for numeric parameters). Types can be atomic (\eg integers and strings) or structured (\ie compound objects). For instance, the \textit{register new user} operation expects a request body containing a JSON object with three string fields: \texttt{username}, \texttt{password}, and \texttt{email}. Similarly, the response schema of the \textit{retrieve all users} operation is a JSON array of JSON objects, each one composed by four fields: three string fields \texttt{username}, \texttt{password}, and \texttt{email}; and a boolean field \texttt{admin}.

\subsection{Mass Assignment Vulnerability}

Developers usually rely on frameworks to develop their REST API. These frameworks, available for all the major web programming languages, provide a set of features and automations to speed up and simplify the development of a REST API. Among the most adopted frameworks we find Spring (Java), Express.js (JavaScript), FastAPI (Python), Flask (Python), Laravel (PHP), and Slim (PHP).


A common practice supported by REST API frameworks is to use a naming convention to 
map input data in HTTP requests to the back-end data representation inside the REST API, such as input parameters to database columns, when they have the same name.
This feature is typically enabled by default, and it is exploited to speed up the development of REST APIs by simplifying the flow of data.


As stated by the OWASP Foundation, a REST API is said to be vulnerable to \textit{mass assignment} when the value of a resource meant to be read-only can be manipulated by external users, by exploiting a misconfiguration of the automatic parameter binding. 


For example, the \textit{VAmPI} API is built on top of the Flask framework, that automatically maps the JSON \texttt{user} object in an HTTP request to the \texttt{User} Python class, that is eventually mapped to the \texttt{users} database table.

The \texttt{users} table (and \texttt{User} class) contains an \texttt{admin} boolean column (class field) meant to be read-only, i.e., while other properties of a user are meant to be edited by the end-user (\eg the {\tt password} or the {\tt email} address), the {\tt admin} field is supposed not to be changed through this operation, in fact it is not documented among the input parameters in the OpenAPI specification of this operation.

However, if the REST framework is not properly configured (as in the case of the default configuration), then the framework will automatically bind an additional \texttt{admin} HTTP parameter in a \textit{register user} request to the corresponding \texttt{admin} column in the \texttt{users} database table. An attacker could forge a request to the operation \texttt{POST /users/register} meant to register a new user, by injecting an additional \texttt{admin} parameter, that is not mentioned in the OpenAPI specification. The additional parameter \texttt{admin} is automatically linked to the \texttt{admin} column in the \texttt{users} table, so the HTTP parameter value controlled by the attacker will overwrite the legitimate value in the database.

Automatic binding boosts productivity: it dispenses developers from the manual configuration of input/output data and database tables/columns relations. Indeed, it is a common feature, provided by all main REST API frameworks. Nevertheless, blind use of this feature often leads to mass assignment vulnerabilities, also known as {\em auto-binding} vulnerabilities or {\em object injection} vulnerabilities.

REST API frameworks nowadays provide a way to mitigate this vulnerability, by allowing developers to explicitly define which HTTP input parameters are allowed to be mapped to the internal data representation. However, developers should be aware of mass assignment vulnerabilities in order to apply this mitigation correctly. 
As a matter of fact, the mass assignment vulnerability is still widespread and listed among the most common vulnerabilities in REST APIs.

\section{Approach}\label{sec:approach}


The OWASP Foundation provides some guidelines~\cite{owasp-guideline} to detect and mitigate mass assignment vulnerabilities, even in the case when APIs source code is not accessible (\ie from a black-box perspective). These require the identification of HTTP requests that may create or update resources in the back-end, and the identification of sensitive object fields (e.g., an \texttt{isAdmin} attribute), analyzing the responses received by the back-end. Guidelines are meant to be exploited manually from developers, hence they do not provide any hint on how to automatize the process of mass assignment vulnerabilities detection.

Hence, we propose a \emph{security testing} approach to automatically find potential mass assignment vulnerabilities in REST APIs. 
Usually, the source code of a REST API is not (or just partially) available. Furthermore, REST APIs are typically composed by many dynamically allocated distributed components (distinctive of a micro-services architecture), making source code analyses computationally challenging. For these reasons, we adopt a \emph{black-box} perspective, i.e., our approach does not assume source code access. This also implies that our approach is applicable to any REST API, independently by the programming language or framework used for its implementation. The only requirements of our approach are: (i) the availability of an OpenAPI specification, to retrieve operations and their input/output message format; and (ii) black-box HTTP access to the REST API under test, to send test requests and receive responses.


Our strategy to test mass assignment vulnerabilities is composed of these main three steps.
\begin{enumerate}
    \item {\em Identification of read-only attributes.} We parse the OpenAPI specification of the REST API under test to list operations and their input/output attributes. Operations are subject to clustering, to group together those that handle similar data resources. We then compare operations handling similar resources to spot read-only attributes, i.e., those that appear as output of {\em read} operations, but not as input in {\em write} operations. These fields are the candidate for mass assignment vulnerabilities.
    \item {\em Test case generation.} Test scenarios are automatically generated as sequences of requests, with the aim of trying to overwrite read-only attributes. Test cases are generated by computing concrete interactions, starting from abstract testing templates. 
    \item {\em Security oracle.} Test cases are executed on the REST API under test, and their execution is monitored by our \emph{security oracle}. The oracle reveals a successful exploitation of a mass assignment vulnerability when a test case manages to overwrite a read-only parameter.
\end{enumerate}

In the following, we detail all the aforementioned steps.

\subsection{Identification of Read-only Attributes}



The first step towards testing mass assignment vulnerabilities consists in the identification of read-only attributes, according to the OpenAPI specification, since they are potential targets of successful attacks. In particular, this requires to identify whether operations read or write data (i.e., infer operations CRUD semantics) and when these actions are performed on the same resource. 




{\bf CRUD semantics and operations resources.} We propose an automatic procedure to infer operations CRUD semantics and the resources that operations act on. This information is collected by inspecting the OpenAPI specification only.

{\em Inferring operations CRUD semantics.} To infer the read/write semantics of operations, we rely on their HTTP method as follows. We assume that {\tt POST} methods are used to \emph{create} (that we annotate as $C$) resources, while \texttt{GET} methods are used to \emph{read} resources. We distinguish reading a single resource (that we annotate as $R$) when the return schema type is a single object, from reading multiple resources at once, that we call \emph{read-multi} (that we annotate as $RM$), when the schema return type is an array of objects. Furthermore, \texttt{PUT} and \texttt{PATCH} methods are considered to be used to update (that we annotate as $U$) resources, while \texttt{DELETE} methods (that we annotate as $D$) are considered to be used to delete resources. 

With respect to the example in Fig.~\ref{code:openapiexample}, the operation \texttt{GET /users} is tagged as a read-multi ($RM$), in fact its return type is an array, while the operation \texttt{POST /users/register} is recognized as create~($C$).

Despite there in no constraint for REST APIs to meet this mapping, typically programmers follow this convention: the converse is generally considered an antipattern and a bad practice~\cite{palma2014detection}.

{\em Grouping operations by resource type.} To detect read-only attributes it is important to understand when operations handle the same type of resource, either in read or in write mode. In fact, as in the example in Fig.~\ref{code:openapiexample}, operations definition does not mention that operations handle {\em user} resources. Moreover, operations handling the same type of resource rarely operate on exactly the same set of fields. For instance, {\em register new user} has three input parameters, while {\em retrieve all users} has four output parameters. 

In order to group together operations with the most {\em similar} parameters, we use clustering. We collect the names of the input/output parameters for all the operations in the OpenAPI specification. These names are normalized using the Porter stemming algorithm~\cite{willett2006porter}, to keep only the root of words (e.g., in case singular/plural is used). After this, duplicate names are discarded and the remaining ones are sorted alphabetically into a global list. Eventually, as shown in Table~\ref{tab:boolean-encoding}, each operation is encoded as a boolean array, with the $i$-th element of the array set to \texttt{true}, when the $i$-th parameter in the global list is an input or output parameter of this operation, according to the OpenAPI specification.

\begin{table}[]
    \centering
	\caption{Boolean encoding of operation parameters for clustering.}
	\label{tab:boolean-encoding}
    \begin{tabular}{l|c|c|c|c}
    \toprule 
    \textbf{Operation}          & \texttt{\textbf{admin}} & \texttt{\textbf{email}} & \texttt{\textbf{password}} & \texttt{\textbf{usernam}} \\ \midrule
    \textit{register user}      & \texttt{False}              & \texttt{True}              & \texttt{True}                 & \texttt{True}                \\
    \textit{retrieve all users} & \texttt{True}              & \texttt{True}              & \texttt{True}                 & \texttt{True} \\    
    \bottomrule           
    \end{tabular}
\end{table}

Then, we apply the expectation maximization (EM) clustering algorithm on this operation representation. The EM clustering algorithm automatically determines the optimal number of clusters, and which operations belong to each cluster, based on common parameters. In our example, it is very likely that the operations {\em retrieve all users} and {\em register new user} would be assigned to the same cluster of {\em user} resources, because they have a very similar set of input/output parameters.

{\em Detecting resource identifiers.} Finally, it is not enough to understand when operations insist on the same resource type (e.g., on {\em users}), we should also detect when the same resource instance is used in subsequent operations in a test scenario (e.g., the same {\em user} resource). To achieve this objective, we detect parameters that act as resource (unique) identifiers, that we call the {\em resource-ids}, commonly used by REST APIs.

We recognize the resource-id with a simple but effective heuristic, that relies on common naming practices\footnote{\url{https://restfulapi.net/resource-naming/}}. Among the input/output parameters in an operation, we consider as resource-id the field that ends with the suffix {\em id} or {\em name}.
In case more than one field satisfies this simple pattern, in a complex schema with different level of nesting, we pick the less nested field. In case we find more than one field at the same level, we prefer the one with {\em id} as suffix. In case the selection is still not unique, we take the first one.

In our example, the field {\tt username} would be selected as resource-id for all the operations on the resource {\em user}. 

{\bf Specification annotation.} Once CRUD semantics, resource types and resource-ids are identified, they are saved in the OpenAPI itself. To this aim, we elaborated an annotation syntax to add these information to the \openapi specification. Annotations are based on custom extension fields as follows.
\begin{itemize}
    \item \texttt{x-crudOperationSemantics} annotates an operation with its CRUD semantics. This field can be \texttt{create}, \texttt{read}, \texttt{update}, \texttt{delete}, \texttt{read-multi}, or \texttt{none}\footnote{The annotation \texttt{none} indicates that no CRUD semantics is present for the operation, since it does not involve any resource (e.g., login operations).}.
    \item \texttt{x-crudResourceType} annotates an operation with the type of the handled resource. This annotation accept a string value.
    \item \texttt{x-crudResourceIdentifier} annotates the parameter that is used as resource-id in the operation.
\end{itemize}

Alternatively, a developer might manually annotate the OpenAPI with these custom tags to skip our automated detection. In fact, despite guidelines and best practices inspired our identifications of the CRUD semantics, resource types and resource-ids detection might be imprecise.

{\bf Read-only parameters detection.} 
After all the required information is collected, we can start the actual identification of read-only parameters.
Inspecting the tags in the instrumented specification (either by means of the clustering or manually), operations that handle resources of the same type are grouped together. 
Read-only parameters are simply detected as those parameters that are available as output in read ($R$, or $RM$) operations, but are never used as input in create ($C$) nor update ($U$) operation in the same group.

In our example, the {\tt admin} parameter would be classified as read-only, because it can be accessed by the read operation {\tt GET /users}, but it can never be modified by the write operation {\tt POST /users/register}.


\subsection{Test Case Generation}

After read-only fields are identified, we need to generate execution scenarios that indeed attempt at overwriting them. 
%
%
The testing approach relies on \emph{abstract test templates} that adopt the following notation.
\begin{itemize}
    \item For a resource of type $\tau$, requests to operations involving $\tau$ resources are labeled as $C_\tau$, $R_\tau$, $RM_\tau$, $U_\tau$, and $D_\tau$, to represent, respectively, requests meant to create, read, read multiple resources, update or delete a resource of type $\tau$.
    \item An operation request is labeled with the plus symbol $C_\tau^{+f}$, when a request to the operation $C_\tau$ is modified by adding the extra parameter $f$ that is not documented in the specification of $C$, in the attempt of overwriting the read-only attribute $f$.
    \item A sequence of requests is tagged with the question mark when it is optional (but recommended). They can be used, for instance, to clean up the REST API from testing data.
\end{itemize}

An abstract test template is formalized as a sequence of labels describing operations and resources type. In particular, we identified two abstract test templates: \emph{Update-injection Sequence} and \emph{Create-injection Sequence}.

\vspace*{5pt}

\noindent{\bf Template 1: Update-injection Sequence}
\begin{center}
    $<C_\tau, R_\tau, U_\tau^{+f}, R_\tau, ( D_\tau, R_\tau )^{?}>$     
\end{center}


This sequence is intended to test a potential mass assignment vulnerability in the update operation $U_\tau$. The sequence starts with the creation of a resource in the REST API. A read operation on this resource is then performed to check a successful creation. The third operation is the update operation $U_\tau^{+f}$ in which the read-only parameter $f$ is injected. The subsequent read operation $R_\tau$ is meant to check if the exploit is successful, by comparing the value of $f$ before and after the injection $U_\tau^{+f}$. In case the read-only parameter has changed, a mass assignment vulnerability has been exposed.

Additionally, the created resource may be deleted, with the aim of cleaning testing side effects (i.e., testing data). In this case, a last read operation checks if the resource has been successfully deleted or if it is still present (i.e., the delete operation has failed).

\vspace*{5pt}

\noindent{\bf Template 2: Create-injection Sequence}
\begin{center}
    $< C_\tau^{+f}, R_\tau, ( D_\tau, R_\tau )^{?}>$
\end{center}

This sequence is intended to test a potential mass assignment vulnerability in the create operation $C_\tau$. The first operation in the sequence tries to create a resource by specifying an additional read-only parameter. Subsequently, a read operation checks if the read-only parameter in the new fresh resource has been successfully overwritten.

As in the previous case, two optional requests close the sequence to clean the state of the REST API: the deletion of the testing resource and a check for successful deletion.

Note that, the last two requests $( D_\tau, R_\tau )^{?}$ in both abstract test templates are not directly involved in the mass assignment vulnerability, indeed they are optional. For instance, in case no delete operation $D_\tau$ is available in the REST API under test, this optional cleanup sequence has to be skipped.



{\bf From abstract to concrete test cases.} 
%
%
Abstract template sequences have to be instantiated in order to be executed on the REST API under test. For each abstract operation in the sequence, we have to generate the corresponding HTTP request, including values for the input parameters.

{\em Input values.}
Values for input parameters of requests (that are not resource-ids) are obtained either from the \openapi specification, or by random generation. The \openapi specification provides for each parameter a set of example values and the parameter default value. Moreover, in the case of enum parameters, also a set containing the supported enum values is usually provided. We resort to these values, when available. In case these values are not available or do not allow generating successful requests (e.g., incorrect example values), we switch to random values, that are  generated to match the parameter format defined in the specification. For string parameters  that specify no specific format, we try to infer the format from the parameter name. For instance, if a field is named \texttt{email}, we try to generate a valid e-mail address. We support more than 20 string formats including dates, timestamps, e-mail addresses, phone numbers, URLs, UUIDs, etc.

{\em Resource-id values.}
All the operations in a test template are supposed to operate on the same resource instance. For example, after creating a resource, the subsequent {\em read} operation is supposed to access the just created resource. To make sure that this is the case, the resource-id field is inspected in the output of {\em create} operations $C_\tau$ and used as input in subsequent 
{\em read}, {\em update} and {\em delete} operations $R_\tau$, $U_\tau$ and $D_\tau$. Since the resource-id would be the same for all the operations in a sequence, in order to keep the notation simple, we did not specify which resource-id is handled by each operation. 

However, sometimes the output of a {\em create} operation $C_\tau$ does not contain the resource-id of the fresh resource, supposed to be used in the subsequent {\em read} operation $R_\tau$. To solve this problem, whenever available, we resort to a read-multi operations $RM_\tau$ to list all the resources of type $\tau$ before and after the creation $C_\tau$. We expect to identify the newly created resource as the difference between these two lists. The two abstract test templates are modified in this way:
\begin{flalign*}
    &\text{Template 1': } <RM_\tau, C_\tau, RM_\tau, 
    U_\tau^{+f}, R_\tau, ( D_\tau, R_\tau )^{?}> \\
    &\text{Template 2': } <RM_\tau, C_\tau^{+f}, RM_\tau, 
    ( D_\tau, R_\tau )^{?}>     
\end{flalign*}

\textit{Concrete test cases.}
Instantiating a concrete test case starting from an abstract template requires to successfully execute each operation from the corresponding sequence. However, there could be multiple candidate operations for a step in the sequence, e.g., more than one operations $R_\tau$ to read a resource of type $\tau$.
A candidate operation can be tested with different input values until it succeeds (status code \twoxx). The maximum number of attempts for testing an operation (\texttt{MAX\_OPERATION\_ATTEMPTS}) is configurable. In case no attempt can obtain a successful response, we consider the whole instantiation of the template as failed. The instantiation of the template is restarted from scratch until all the operations are executed successfully, for a maximum number of attempts \texttt{MAX\_TEMPLATE\_ATTEMPTS}. 

A larger number of attempts will increase the probability of successfully instantiating a sequence, but also a longer test generation time.

\subsection{Security Oracle}

The responsibility of the security oracle is to classify test case executions and judge whether they are able to reveal a mass assignment vulnerability in the REST API under test. To this aim, the oracle verifies that the same resource-id is used by all the operations in a sequence, and that the injection was successful with a value different from the default one. These three aspects of the oracle are described in the following.

{\em Same resource-id check.} As described earlier, all the operations in a test scenario should, by construction, handle the same resource instance, by using the same value of the resource-id field. The security oracle has to check that this constraint is met when a test case is executed.

{\em Read-only fields overwrite.} When the oracle detects that a read-only value has been successfully overwritten, it reports a mass assignment vulnerability. In practice, the oracle checks if the injected update request $U_\tau^{+f}$, or the injected create request $C_\tau^{+f}$ could overwrite the value of the read-only parameter $f$. This is achieved by comparing the injected value of the read-only parameter $f$ with the value observed from the subsequent read request $R_\tau$. In case the two values correspond, the test case managed to write the read-only field $f$.

{\em Injection of default values.} 
When instantiating an abstract test template into a concrete test case, values should be chosen for all the input fields. When choosing the value for the read-only parameter, we could unintentionally guess its {\em default} value, especially for boolean or enum types that support a limited set of possible values. If the subsequent read operation reveals that the resource contain the injected value in the read-only field, our oracle could report a vulnerability. However, it could be the case that the attack failed, and the read-only parameter has the value that it would have had in the nominal case, i.e., its default value. To check for this case and avoid a false positive, each test case that our oracle classifies as a successful injection is repeated using a different value for the injected read-only parameter. Our objective is to make sure that the injection is successful with at least two distinct values of the read-only field, and avoid false positives due to default values.

{\bf Identification of other defects.}
Despite the objective of this oracle is to reveal mass assignment vulnerabilities when running test cases, it is also able to detect other kind of defects.

Injection consists of a request that violates the specification by adding an undocumented field. Thus, a correctly implemented REST API should respond with a \fourxx status code, and reject the malformed request with the unexpected parameter $f$, because it is not consistent with the documented request format. A \twoxx status code, instead, means that the REST API accepted as valid the injected, and thus invalid, request. Even if injection did not happen, accepting a malformed request by itself is a defect. Additionally, a response status code \fivexx stands for an internal server error that was not handled correctly (e.g., because of an uncontrolled exception). This suggests a potential implementation defect, that is however different from the mass assignment vulnerability.

\section{Experimental Setting}
\label{sec:settings}

In this section, we present the experimental setting for the empirical validation of our approach.

\subsection{Research Questions}

First, our approach requires to identify operations CRUD semantics, what resource types they handle, and what parameter they use as resource-id. Hence, our empirical investigation first assesses the accuracy of the automated identification of such information.

\begin{description}
	\item[\textbf{RQ1:}] What is the accuracy of the automated identification of operations CRUD semantics, resource types, and resource-id parameters?
\end{description}

Second, the main objective of our approach is to reveal mass assignment vulnerabilities in REST APIs, this is investigated by the following research question.

\begin{description}
	\item[\textbf{RQ2:}] What is the accuracy in revealing mass assignment vulnerabilities in REST APIs?
\end{description}

Finally, our approach should be able to deal with real-world production-ready REST services, that can be complex and large in size. Hence, the last research question investigates the scalability of the approach on large REST APIs.
\begin{description}
	\item[\textbf{RQ3:}] Does the proposed approach to detect mass assignment vulnerabilities scale to large REST APIs?
\end{description}

\subsection{Case Studies}
\label{sec:case-studies}

The experimental validation is conducted on a set of REST APIs. However, considering that test cases are supposed to exploit a security vulnerability, it would not be ethical to test publicly hosted APIs (such as those on \emph{APIs.guru}\footnote{\url{https://apis.guru/browse-apis/}}), since their integrity could be compromised by a successful attack. Hence, we opted for case studies that we can download and run in a controlled environment.


For these reasons, we looked for open-source projects hosted on \emph{GitHub}, so that we can compile and install them locally. This also grant us full control over the REST APIs state and data, allowing us to restore an API initial state and, hence, avoiding potential side effects originated by previous injections.


We queried the GitHub search engine with the following base strings: ``REST'', ``RESTful API'', ``mass assignment'', ``autobinding'', ``object injection'', ``\openapi'' and ``Swagger''. Then, we also added query strings representing framework commonly used to implement REST APIs, such as ``swagger-ui'', ``SpringFox'', ``swagger-jsdoc'', ``flask-swagger''. Among the REST APIs resulting from the search, we selected those containing an \openapi specification, which is a requirement of our black-box testing approach.

These potential case studies have been downloaded, compiled and run to discard those that failed either in compiling or in running. 
After this last filtering, our final set of case studies is shown in Table~\ref{tab:csSummary}. It consists of 5 REST APIs, written in different programming languages (Java, ASP.NET, Python, and Node.js), based on different frameworks and DBMSs, for a total of 26 endpoints and 47 operations. 

The table also reports the number of vulnerabilities in each case studies. 
While {\em VAmPI} and {\em OWASP} already contained some vulnerabilities (1 and 4, respectively), the other case studies did not, so they have been manually seeded with vulnerabilities with the following procedure. 
We edited the internal representation of a resource in the REST API (e.g., resource {\em user}) to add a new field with arbitrary name. Then, we edited the code of a {\em write} operation that handles this type of resource (e.g., {\tt PUT /user}) to enable automatic mapping. Eventually, we edited a {\em read} operation for this resource type and its section in the OpenAPI specification to add the novel field to those that can be read.

Moreover, we also prepared non-vulnerable (safe) versions for all the case studies. \textit{VAmPI} supports an environment boolean variable to enable or disable the vulnerability. The repository of \textit{OWASP} contains both a vulnerable and a patched version. For the remaining case studies, we fixed the vulnerable version, adopting the mitigation features of frameworks.





\begin{table}[tb]
	\centering
	\caption{Vulnerable REST APIs considered for assessing the accuracy of the approach.}
	\label{tab:csSummary}
	\begin{tabular}{l|c|c|c|c}
		\toprule
		\textbf{Case study} & \textbf{Language} & \textbf{Framework} & \textbf{\# Ops.} & \textbf{\# Vuln.} \\
		\midrule
		VAmPI~\cite{cs-vampi} & Python & Flask & 12 & 1 \\
		OWASP~\cite{cs-owasp} & Java & Spring & 10 & 4 \\
		Toggle~\cite{cs-toggle} & ASP.NET & .NET Core & 16 & 2 \\
		Bookstore~\cite{cs-bookstore} & Java & Spring & 5 & 1 \\
		CRUD~\cite{cs-crud} & Node.js & Express & 4 & 2 \\
		\bottomrule
	\end{tabular}
\end{table}

Eventually, our approach requires a correct \openapi specification as input. However, in some cases, specifications of open-source REST APIs are incorrect or incomplete. So, before running the experiment, all the \openapi specifications have been checked against the source code by a team of experts, including an author of this paper and two other external contributors. Few corrections have been applied, \eg response schemas have been added in case they were missing. The same team also elaborated the ground truth for CRUD semantics, resource types, and resource-ids of all the operations.

Finally, to answer the third research question, we considered a collection of more complex case studies, namely 10 mainstream Google REST APIs (Gmail, Google Analytics, Calendar, Classroom, Custom Search, Drive, Fitness, My Business, Search Console and YouTube).





\subsection{Experimental Procedure}

We implemented our approach as a testing strategy, \texttt{MassAssignmentSecurityTestingStrategy}, on top of RestTestGen~\cite{Corradini2022}. RestTestGen is an open-source REST APIs testing tool (available on GitHub\footnote{\url{https://github.com/SeUniVr/RestTestGen}}, where you can also find the mass assignment strategy) and it provides some commodity and utility features that facilitated the implementation of our approach. They are a solid parser for the \openapi specification, procedures to construct and send HTTP requests to the REST API under test, and receive and parse the corresponding HTTP responses.

We configured the parameters of the template instantiation with the following values: \texttt{MAX\_OPERATION\_ATTEMPTS = 12}, and \texttt{MAX\_TEMPLATE\_ATTEMPTS = 3}. These values happened to be an optimal trade-off between testing time and successful vulnerabilities identifications during our dry run experiments.

Case studies have been taken into a testable state, \ie	their states have been initialized with some data, preparatory for testing. Most of the case studies  come with a specific script, while some other required several manual HTTP interaction in order to provide initialization data.

Our testing strategy has been executed on each case study, both on the vulnerable and safe version. The process has been repeated ten times, to control potential random variation due to the non-deterministic algorithms in our approach. After each testing session, we restored the same identical state of the REST APIs under test.





\subsection{Metrics}

In our experiment, we computed and collected the following metrics.


	{\bf Correctness of CRUD identification.} For each REST API, we compute the ratio of the operations for which the correct CRUD semantic was identified divided by total number of the operations with a CRUD semantics in the specifications, according to the manually provided ground truth.

	{\bf Clustering correctness.} We manually inspect the result of clustering to identify what is the resource type that is handled by the majority of the operations in each cluster. Those operations that handle a resource type different from the majority do, are considered as assigned to the wrong cluster. This metric is then computed as the ratio of the operations assigned to the correct cluster divided by the total number of CRUD operations. 

	{\bf Correctness of resource-id identification.} This metric is computed as the ratio of the operation schemas for which the resource-id field has been correctly identified divided by the total number of schemas of CRUD operations that contain a resource-id. Operations can have an input schema and/or an output schema.

	{\bf Accuracy in vulnerability detection.} To measure the accuracy in detecting mass assignment vulnerabilities, we use standard information retrieval metrics.
	\begin{itemize}
	\item {\em True Positives (TP):} number of vulnerabilities that are correctly detected.
	\item {\em False Positives (FP):} number of cases reported as vulnerabilities that are not actually vulnerable (false alarms).
	\item {\em False Negatives (FN):} number of vulnerabilities that are missed (not reported by the tool).
	\item {\em Precision (Pr):} number of correctly detected vulnerabilities on the number of reported vulnerability (\emph{Pr} = \emph{TP} / (\emph{TP} + \emph{FP})).
	\item {\em Recall (Re): } number of correctly reported vulnerabilities on the number of vulnerabilities (\emph{Re} = \emph{TP} / (\emph{TP} + \emph{FN})).
	\end{itemize}
	





\subsection{Threats to Validity}

Here we discuss the main threats to the validity of our findings.

Threats to the {\em conclusion validity} are concerned with issues that affect the ability to draw the correct conclusion. To limit this threat, we considered standard information retrieval metrics (i.e., precision and recall). Additionally, to increase heterogeneity of samples in the data set, we considered software projects written in multiple programming languages and based on different frameworks.

Threats to {\em internal validity} concern the subjective factors that might have affected the results, such as the definition of the gold standard to compare the results of our approach. To limit this subjectivity, we inspected the source code of operation to determine its actual CRUD semantic, the type of the read/written resource and what field is used as resource-id (if any). 

The threats to {\em construct validity} concern the data collection and analysis procedures. To mitigate this threat, we made sure that, when injecting faults in case studies, the vulnerability definition as explained by the OWASP Foundation was precisely followed. 

Threats to {\em external validity} concern the generalization of our findings. Despite our analysis considers REST APIs written in different programming languages and based on different rest frameworks, our analysis could still be biased because we only considered open-source projects. Our results might not extend in general to other software projects, for instance to closed-source industrial software. Only replications with more samples can confirm our findings beyond our experimental settings.

\section{Experimental Results}
\label{sec:results}

This section presents the results of our experimental validation. The replication package for this experiment is available online~\cite{ReplicationPackage}.


\subsection{RQ1: Accuracy of CRUD Semantics, Resource Type, and Resource-id}

Before starting the actual test case generation, some preliminary results should be computed purely based on the OpenAPI specification. The accuracy of this first step is investigated by the first research question, for all the case studies. The tool automated labeling has been compared with the gold standard, and results are reported in Table~\ref{tab:rq1}. For each case study (first column), the table reports the correctness of the CRUD semantics identification (second column), the clustering correctness (third column), and the correctness of resource-id identification (fourth column).

\begin{table}[t]
	\centering
    \caption{Correctness of CRUD semantics identification, of operations clustering, and resource-id identification.}
    \label{tab:rq1}
    \begin{tabular}{l|r|r|r}
    \toprule
    \textbf{Case study} & \textbf{CRUD} & \textbf{Clustering} & \textbf{Resource-id} \\ \midrule
    VAmPI  & 100\% & 100\% & 67\% \\ 
    OWASP  & 100\% & 80\% & 100\% \\ 
    Toggle  & 88\% & 88\% & 100\% \\ 
    Bookstore & 100\% & 100\%  & 100\% \\ 
    CRUD  & 100\% & 100\% & 100\% \\  
    \midrule
    \textbf{Average} & 98\% & 94\% & 93\% \\ 
    \bottomrule
    \end{tabular}
\end{table}

The identification of CRUD semantics is 100\% for almost all the case studies, reaching an average correctness of 98\%. Such a high accuracy is due to a disciplined use of HTTP methods in the observed REST APIs. Despite accuracy was still very high (i.e., 88\%), a wrong semantics have been identified in some operations of {\em Toggle}. A manual investigation revealed that {\em read multi} operations were automatically labeled as {\em read}, because of an uncommon return type. Instead of returning an array of objects, a single JSON object is returned, that, in turn, contains many sub-objects that corresponds to resources.

The clustering correctness is also quite high, with an average value of 94\%. Operations that have been allocated to the wrong cluster turned out to be only {\em delete} operations. This is due to the fact that, typically, {\em delete} operations only require no other input than the resource-id of the resource to delete. When the other operations insisting on the same resource type have a lot more input parameters than the \textit{delete} operation, the clustering algorithm is very likely to assign the latter to a separate cluster (only containing the \textit{delete} operation itself).
Nevertheless, availability of {\em delete} operations is not strictly required for our testing approach ($D_\tau$ is optional in both {\em Template~1} and {\em Template~2}).

Resource-ids identification is also quite accurate, with an average correctness of 93\%. Inaccurate detection was due to custom naming in some case studies that violated common practices. For instance, in {\em VAmPI} uses the field \texttt{book\_title} as resource identifier, instead of a more intuitive \texttt{book\_id} or \texttt{book\_name}.

\begin{table}[t]
	\centering
	\caption{Accuracy in revealing mass assignment vulnerabilities.}
	\label{tab:rq2}
	\resizebox{\linewidth}{!}{%
		\begin{tabular}{l||c|c||c|ccc|cc}
			\toprule
			\textbf{Case study} & \multicolumn{2}{c||}{{\bf Safe}} & \multicolumn{6}{c}{{\bf Vulnerable}}\\ \midrule
			& \textbf{Tests} & \textbf{FP}  & \textbf{Tests} & \textbf{TP} & \textbf{FP} & \textbf{FN} & \textbf{Pr} & \textbf{Re} \\ \midrule
			VAmPI & 4.0 & 0.0 & 4.0 & 1.0 & 0.0 & 0.0 & 100\% & 100\% \\ 
			OWASP & 8.0 & 0.0 & 7.4 & 3.6 & 0.0 & 0.4 & 100\% & 90\% \\ 
			Toggle & 2.0 & 0.0 & 2.0 & 2.0 & 0.0 & 0.0 & 100\% & 100\% \\ 
			Bookstore & 2.0 & 0.0 & 2.0 & 1.0 & 0.0 & 0.0 & 100\% & 100\% \\
			CRUD & 2.0 & 0.0 & 2.0 & 2.0 & 0.0 & 0.0 & 100\% & 100\% \\ \bottomrule
		\end{tabular}
	}
\end{table}


Based on this results we can answer RQ1 in this way:
\begin{center}
\fbox{
\begin{minipage}[t]{0.9\linewidth}
\textit{Our approach shown a quite high accuracy in analyzing the OpenAPI specification, because the CRUD semantics was detected correctly for 98\% of the operations, 94\% of the operations have been grouped together when they handle a resource of the same type, and the 93\% of resource-ids have been correctly identified.}
\end{minipage}
}
\end{center}

\subsection{RQ2: Accuracy of Vulnerability Detection}

Then, we applied our testing approach to two variants of each case study, i.e., with and without vulnerabilities. The outcome of the oracle evaluating the automatically generated test cases has been compared with the gold standard. The accuracy of vulnerability detection is shown in Table~\ref{tab:rq2} in terms of true positives (TP), false positives (FP), false negatives (FN), precision (Pr), and recall (Re). The table also reports the number of concrete test cases that could be successfully generated for the read-only fields detected on each variant of each case study. 
The table reports the average values for the ten repetitions of each testing session (thus, decimal values).



Accuracy in testing mass assignment vulnerabilities shown to be very high. All the vulnerabilities are detected in almost of all case studies, with no false positive. Only in one run out of ten (this is why the average TP is a decimal number in the table) on \textit{OWASP}, our approach missed few vulnerabilities. Manual investigation on test execution traces revealed that, in a particular execution on \textit{OWASP}, our approach could only instantiate two sequences out of eight, so the corresponding vulnerabilities could not be tested. 



As a side effect of security testing, our tests could also reveal the presence of other defects such as crashes (status code \fivexx) in all case studies, except for \textit{Toggle}. In total total, 24 HTTP responses occurred with status code \fiveoo. Manual inspection on HTTP requests and responses revealed that they originated from 5 distinct bugs.

Based on these results, we can formulate the subsequent answer to RQ2:
\begin{center}
\fbox{
\begin{minipage}[t]{0.9\linewidth}
\textit{The accuracy in revealing mass assignment vulnerabilities with automated black-box testing is very high, because all the vulnerabilities could be detected in our case studies with no false positives, with the only exception of a single execution out of ten in a single case study.}
\end{minipage}
}
\end{center}








\subsection{RQ3: Scalability of the Approach}
To study the scalability of our approach, we ran the static analysis part on the specifications of 10 mainstream Google services, listed in Table~\ref{tab:rq3}, that include a total of 454 operations. To mitigate the impact of non-determinism introduced by clustering, each service have been tested 5 times independently. Note that, we could not run the dynamic test case generation part for ethical reasons, i.e., we meant to avoid the risk of mounting a successful injection attack on a service that runs in production. 

The static analysis of all services took 37.2 seconds (on average) and, in total, it revealed 981 read-only fields. Even if we could not run the automated test case generation part, we expect that not a long time would be needed to validate these candidate vulnerabilities, with few test cases. So, we can speculate that scalability on large REST APIs does not represent an issue for our approach.


Thus, we can answer RQ3 in this way:
\begin{center}
\fbox{
\begin{minipage}[t]{0.9\linewidth}
\textit{Our approach seems to scale well on large APIs, because it took 37.2 seconds to analyze remarkably large OpenAPI specifications, detecting 981 read-only parameters (on a total of 454 operations) that are potential candidates of mass-assignment vulnerabilities.}
\end{minipage}
}
\end{center}

\begin{table}[tb]
	\centering
	\caption{Google REST APIs considered for assessing the scalability of the approach.}
	\label{tab:rq3}
	\begin{tabular}{l|c|c|c}
		\toprule
		\textbf{Case study} & \textbf{\# Ops.} & \textbf{Time (s)} & \textbf{\# Read-only fields} \\ \midrule
		Gmail  & 68 & 3.0 & 23 \\ 
		Analytics  & 88 & 5.0 & 166 \\ 
		Calendar  & 37 & 2.0 & 11 \\ 
		Classroom & 61 & 5.0 & 15 \\ 
		Custom Search & 2 & 1.0 & 66 \\  
		Drive & 48 & 3.0 & 49 \\ 
		Fitness & 13 & 1.4 & 4 \\ 
		My Business & 50 & 7.4 & 527 \\ 
		Search Console & 11 & 1.0 & 10 \\ 
		YouTube & 76 & 8.4 & 110 \\   
		\midrule
		\textbf{Total} & \textbf{454} & \textbf{37.2} & \textbf{981} \\ 
		\bottomrule
	\end{tabular}
\end{table}

\section{Related Work}
\label{sec:related}




While most of the literature on automated testing on REST APIs is focused on functional testing, just few approaches to automate security testing are starting to rise~\cite{Mai2020MetamorphicSecurityTesting,Atlidakis2020SecurityProperties,Yang2017RestSep,Yang2016RestPL}, in order to detect potential security vulnerabilities in REST APIs. 

In particular, Mai et al.~\cite{Mai2020MetamorphicSecurityTesting} use metamorphic relations to address the oracle problem: 22 system-agnostic metamorphic relations are defined in order to automate security testing in Web systems. Their approach is not black-box and does not target REST APIs. Furthermore, mass assignment vulnerabilities are not part of their catalog of metamorphic relations.

Atlidakis et al.~\cite{Atlidakis2020SecurityProperties} introduce four security rules that capture desirable properties of REST APIs, specifically meant to test some security relevant aspects of Azure REST APIs. In particular, their objective is to test four conditions: (i) use-after-free, when it is possible to read a resource after it has been deleted; (ii) resource-leak, when a resource can be read after its creation failed; (iii) resource-hierarchy, when a resource is accessible under the wrong parent resource; and (iv) user-namespace, when a private resource of a user is accessible to another user. Despite being effective, their approach targets particular security aspects, that do not comprise mass assignment vulnerabilities. Recall that, the latter is indicated by the OWASP Foundation as one of the most important REST APIs vulnerability to mitigate.

Luo et al.~\cite{Yang2017RestSep,Yang2016RestPL} focus on access control. In their first work~\cite{Yang2017RestSep}, their aim is to simplify the privilege partitioning problem into a classification problem of RESTful functions. They propose a REST API classification approach (RestSep) based on genetic algorithms. In their second work~\cite{Yang2016RestPL}, they propose a policy language (RestPL) to express authorization policies for REST APIs. A RestPL policy can be automatically generated from an actual request, which helps to mitigate users intervention. Access control is surely an import aspect that testing tools should address, but it is not related to a mass assignment defect. 

To the best of our knowledge, no approach is available to automatically test REST APIs with respect to mass assignment vulnerabilities. 
%
%
%

Remaining literature on functional testing is mainly split into two different lines of work. 

One consists in \emph{white-box} approaches, that rely on the availability of REST APIs source code to perform static analysis, or to instrument it to collect execution traces and metric values. In this context, Arcuri~\cite{Arcuri2019EvoMaster} proposes a fully automated solution to generate test cases with evolutionary algorithms, that requires the \openapi specification and the access to the Java bytecode of the REST API to test. This approach has been implemented as a tool prototype called EvoMaster, extended with the introduction of a series of novel testability transformations aimed at providing guidance in the context of commonly used API calls~\cite{Arcuri2020TestTransformations}. Unfortunately, white-box approaches are very often not practically usable in the context of REST APIs, where usually the source code is not available.

On a complementary direction, \emph{black-box} approaches do not require any source code, which is often the case when using closed-source components and libraries. \emph{Fuzzers}~\cite{APIFuzzer,FuzzLightyear,FuzzySwagger,SwaggerFuzzer,TnTFuzzer} are black-box testing tools that generate new tests starting from previously recorded API traffic: they fuzz and replay new traffic in order to find bugs. Some of these also exploit the \openapi specification of the service under test~\cite{FuzzLightyear,FuzzySwagger,SwaggerFuzzer}.

Godefroid et al.~\cite{Godefroid2020Fuzzing} propose a methodology to fuzz body payloads intelligently using JSON body schemas and advanced fuzzing rules. 
Even if they are automatic black-box tools, their goal is to generate input values to tests, so they cannot be used as standalone testing tools (except for the approach of Godefroid et al.~\cite{Godefroid2020Fuzzing} that has been implemented in \rler).

Ed-douibi et al.~\cite{EdDouibi2018AutomaticTestsGen} propose a model-based approach for black-box automatic test case generation of REST APIs. A model is extracted from the \openapi specification of a REST API, to generate both nominal test cases (with input values that match the model) and faulty test cases (with input values that violate the model). 

Karlsson et al.~\cite{Karlsson2020QuickREST} propose QuickREST, a tool for property-based testing of RESTful APIs. Starting from the \openapi specification, they generate test cases with the aim to verify whether the API under test complies with some properties (\ie definitions) documented in the specification (\eg status codes, schemas). 

Segura et al.~\cite{Segura2018MetamorphicRelations} propose another black-box approach for REST APIs testing, with an oracle based on (metamorphic) relations among requests and responses. For instance, they send two queries to the same REST API, where the second query has stricter conditions than the first one (\eg by adding a constraint). The result of the second query should be a proper subset of entries in the result of the first query. When the result is not a sub-set, the oracle reveals a defect. However, this approach only works for search-oriented APIs. Moreover, this technique is only partially automatic, because the user is supposed to manually identify the metamorphic relation to exploit, and what input parameters to test.

Corradini et al.~\cite{Corradini2022} propose RestTestGen, an automated black-box test case generation tool for REST APIs. The approach allow to test nominal and error scenarios, and the test generation strategy is based on the Operation Dependency Graph, a graph which encodes data dependencies among the operations available in the API. 

Atlidakis et al.~\cite{Atlidakis2019RESTler} propose RESTler, a stateful REST API fuzzer developed at Microsoft Research. RESTler generates stateful sequences of requests by inferring producer-consumer relations between request types described in the specification. It also dynamically analyzes responses to intelligently build request sequences and avoiding sequences leading to errors.

Martin-Lopez et al.~\cite{MartinLopez2020RESTest} propose RESTest, another automated black-box testing tool for RESTful APIs. The peculiarity of this tool is the inter-parameter dependencies support. Indeed, some REST APIs impose constraints restricting not only input values, but also the way in which input values can be combined to fill valid requests.

Finally, Laranjeiro et al.~\cite{Laranjeiro2021BBOXRT} propose bBOXRT, a black-box robustness testing tool for RESTful APIs. The aim of bBOXRT is to assess the robustness of REST APIs observing the behavior of services under test when providing invalid requests. The tool provide a fault model consisting of 57 different mutations applicable to input parameters of various types (numbers, strings, booleans, dates, times, arrays, etc.).

All the aforementioned black-box testing tools for REST APIs are not meant to spot security vulnerabilities, and mass assignment in particular. Nevertheless, they can be extended in order to implement the methodology presented in the present work. In this respect, we chose to apply our testing strategy on top of RestTestGen, since it provides a modular and easily extensible architecture.

\section{Conclusion}
\label{sec:conclusion}


Dependability and confidentiality of data rely on the correct and secure implementation those REST APIs that are supposed to enforce appropriate data access policies. Mass assignment is among the most common vulnerabilities in REST APIs, often caused by wrong settings in web frameworks. These vulnerabilities might allow an attacker to directly override private internal data structures in the REST API back-end, such as sensitive database columns. 

We proposed an automated approach based on black-box testing to reveal mass assignment vulnerabilities. Starting from the OpenAPI specification, we apply clustering to identify those operations that are very likely to operate on the same resources. Cluster content is compared to identify resource fields that are not supposed to be exposed in write operations, i.e., read-only data according to the developer intention. Abstract test templates are turned into concrete test cases, to test each alleged read-only field according to mass assignment vulnerability. To the best of our knowledge, this is the first work that proposed an automated approach to reveal mass assignment vulnerabilities using automated black box testing.

Experimental results suggest that a variety of programming languages are prone to these vulnerabilities, when web frameworks are not appropriately configured. As future work, we plan to extend our experimental validation to either a more complete set of open-source projects (hence, more programming languages and web frameworks) and closed-source industrial projects from our industrial partners. Additionally, we plan to extend our approach to other kind of programming defects and security vulnerabilities of REST APIs, such as {\em broken object level authorization} and incorrect {\em access control}. Finally, we plan to conduct user studies with developers to identify the most appropriate way to report vulnerabilities in REST APIs, to support a fast and accurate fix of these defects.

\bibliographystyle{IEEEtran}
\bibliography{bib}

\end{document}